\documentclass[aps,twocolumn,showpacs,superscriptaddress,prb]{revtex4}

\usepackage{dcolumn}
\usepackage{bm}
\usepackage{amssymb}
\usepackage{amsmath}
\usepackage{graphicx,epsfig}

\begin{document}
\title{ Superconductor-ferromagnet junction phase qubit }
\author{Tae-Wan Noh}
\affiliation{Department of Physics, Korea Advanced Institute of Science and Technology,
Daejeon 305-701, Korea}
\author{Mun Dae Kim}
\email{mdkim@yonsei.ac.kr}
\affiliation{Institute of Physics and Applied Physics, Yonsei University, Seoul 120-749, Korea}
\affiliation{Korea Institute for Advanced Study, Seoul 130-722, Korea}
\author{H.-S. Sim}
\affiliation{Department of Physics, Korea Advanced Institute of Science and Technology,
Daejeon 305-701, Korea}

\begin{abstract}
We propose a scheme for a phase qubit in an SIFIS junction, consisting of
bulk superconductors (S), a proximity-induced ferromagnet (F), and insulating barriers
(I). The qubit state is constituted by $0$ and $\pi$ phase
states of the junction, in which the charging energy of the junction
leads to the superposition of the two states. The qubit is operated
by the gate voltage applied to the ferromagnet, and insensitive to
the decoherence sources existing in other superconducting qubits. We
discuss a scalable scheme for qubit measurement and tunable
two-qubit coupling.
\end{abstract}


\maketitle

 A superconducting qubit is one of the most
promising candidates for a basic element of quantum computing. There
are different types of the qubit, such as charge qubit
\cite{ChargeQ}, charge-flux qubit \cite{ChargePhase}, flux qubit
\cite{Mooij}, and phase qubit \cite{Berkley}. They are prepared by
manipulating either the charge or the phase degrees of freedom of a
Josephson junction, and operated by gate voltage or by magnetic
microwave pulse. In this work, we propose a new superconducting
qubit, which is based on the phase degree of freedom and operated by
gate voltage. Our qubit is insensitive to the decoherence coming
from the fluctuations of background charges and microwave pulses,
which affect charge qubits and flux qubits, respectively.
Further, the gate voltage operation is suitable for  the individual
addressing in a scalable design of quantum computing and provides a fast
quantum gate operation.

On the other hand, in a superconductor-ferromagnet-superconductor (SFS)
junction, 
the phase difference of $\pi$ appears between the two superconductors under certain
conditions. \cite{Buzdin} The so-called $\pi$ state
is caused by the Zeeman energy induced by the effective magnetic field in the ferromagnet, which
modulates the superconducting pair amplitude.
The ground-state transition of the junction  between the $\pi$ state
and the $0$ state was experimentally observed \cite{Sellier,Kontos},
by tuning temperature or the length of the ferromagnet.
There were proposals \cite{Yamashita,Yamashita06}
in which the $\pi$ state is used for
qubit manipulation in a superconducting ring.
Here the $\pi$ state is {\it not} a qubit state, but
gives rise to an effective magnetic flux in the qubit loop.

In this work, we propose a
phase qubit formed in an SIFIS junction, where a ferromagnet (F)
is sandwiched between two bulk superconductors (S) through
insulating barriers (I). The qubit is constituted by the 0 state
and the $\pi$ state of the junction, when both the phase states
are stable, energetically degenerate, and form a superposition due
to the charging energy of the barriers.
We demonstrate that the qubit can be operated by gate voltage applied to the
ferromagnet, contrary to usual phase qubits.
%
%
We also propose a scalable scheme for tunable and switchable two-qubit
coupling and qubit readout.

\begin{figure}[b]
\vspace{0.cm}
\includegraphics[width=0.35\textwidth]{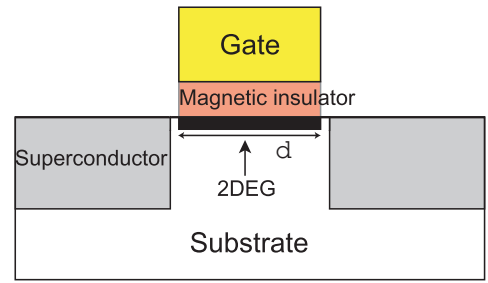}
\vspace{-0.cm}
\caption{(Color online)
SIFIS junction. It consists
of a proximity-induced ferromagnetic 2DEG and two superconductors.
A gate voltage is applied
to the 2DEG to operate the phase qubit formed in the
junction.
The gaps between the 2DEG and the superconductors indicate insulating barriers.
}
\label{fig1}
\end{figure}


The schematic view of our SIFIS junction is shown in Fig.~\ref{fig1}.
The ferromagnetic part is effectively formed by a semiconductor
two-dimensional electron gas (2DEG) underneath a ferromagnetic insulator and a gate voltage $V_g$.
The magnetic field induced by the ferromagnetic insulator causes
the effective ferromagnetic correlation in 2DEG,
the Zeeman energy $h_0$
of the electrons.
And the Fermi wavevector $k_F^{(F)}$ of 2DEG
(thus the phase state of the junction, as shown below) is controlled by $V_g$.



To understand the phase state of the junction, we first ignore the charging energy of the junction.
In this limit, the quasi-particle state $\psi(z)=(u_{\sigma}(z), v_{\bar{\sigma}}(z))^T$
of the SIFIS junction in Fig.~\ref{fig1} can be described by the
Bogoliubov-de Gennes  equation, \cite{BdeG}
\begin{equation}
\left( \begin{array}{cc}
  \!\!\! \ H_0  - \rho _\sigma h(z)  & \Delta (z) \\
  \Delta^* (z) & \!\!\!\!\!-H_0  + \rho _{\bar{\sigma}}h(z)
\end{array}  \right)
\psi(z)=E\psi(z),
\label{BdG}
\end{equation}
where $z$ is the coordinate along the junction direction, $u_\sigma$
($v_{\bar{\sigma}}$) is the electron-like (hole-like) component of
the state, $\sigma$ is the spin index, and $\bar{\sigma} = -
\sigma$. Here $H_0  = -\hbar ^2\nabla^2/2m -\mu$ and $\rho_\sigma = 1 (-1)$ for $\sigma =
\uparrow (\downarrow)$.
The chemical potential $\mu$ equals to
the Fermi energies of the superconductors and the ferromagnetic 2DEG,
$E^{(S)}_F = \hbar ^2 {k^{(S)}_F}^2/2m$ and $E^{(F)}_F =
(E^{(\uparrow)}_F+ E^{(\downarrow)}_F )/2 = \hbar^2
{k^{(F)}_F}^2/2m$.   The Zeeman energy $h(z)$ of electrons
exists only in the
ferromagnetic part, $h(z) = h_0 \theta(z) \theta(d-z)$, where $\theta(z)$
is the step function and $d$ is the length of the 2DEG.
The superconducting pair potential is
approximated to have a step form, $\Delta(z) = \Delta_0 [ \exp (i \phi_L)
\theta(-z) + \exp (i \phi_R) \theta (z-d)]$, which is valid
at low temperature $T \ll T_c$. Here $\phi_{L(R)}$ is the phase of the left (right)
superconductor and $T_c$ is the
critical temperature of the superconductors.

\begin{figure}[b]
\vspace{9cm}
\includegraphics{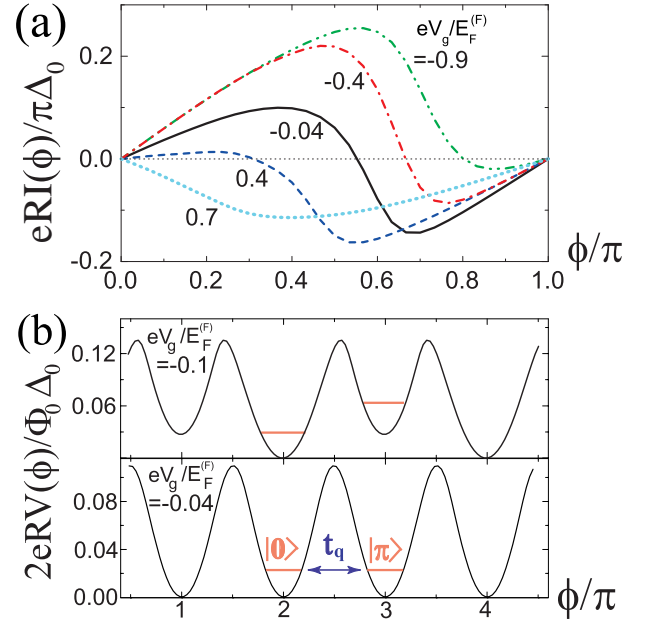}
\vspace{-0.cm}
\caption{(Color online) (a) Current-phase relation $I(\phi)$ at the
vicinity of the crossover ($eV_g/E^{(F)}_F \simeq -0.04$) between $| 0 \rangle$
and $| \pi \rangle$.  (b) Potential energy $V(\phi)$. At $eV_g/E^{(F)}_F \simeq -0.04$,
$|0\rangle$ and $|\pi\rangle$ are degenerate, while the
degeneracy is lifted at $eV_g/E^{(F)}_F=-0.1$. The arrow indicates the
tunneling between $| 0 \rangle$ and $| \pi \rangle$.
In (a) and (b), we choose $T/\Delta_0 = 0.05$, $h_0/E^{(F)}_F = 0.025$,
$R \equiv 2 \pi^2 \hbar /w e^2 {k^{(F)}_F}\approx 200$~$\Omega$, the Fermi
wavelength $\lambda_F=$ 50~nm, and the width of 2DEG $w = 3$~$\mu$m.}
\label{CPR}
\end{figure}

The phase state of the junction can be tuned by $V_g$.
In Fig.~\ref{CPR}(a), we plot the Josephson current of the junction, calculated from the relation, $I(\phi) = (e
\hbar / 2im \beta) \lim_{z\rightarrow z'} (\frac{\partial}{\partial
z'}-\frac{\partial}{\partial z}) \sum_{\omega_n} \textrm{Tr}
[G_{\omega_n}(z,z')]$. Here, $G_{\omega_n}$ is the Green function
obtained from Eq. (\ref{BdG}), $\phi = \phi_L - \phi_R$  is the
phase difference, $\beta = 1 / k_B T$, and $\omega_n$ is the
Matsubara frequency.
In Fig.~\ref{CPR}(b), we plot the potential energy,
$V(\phi) = \frac{\Phi_0}{2\pi} \int^{\phi} _0 I(\phi') d\phi'$,
obtained from $I(\phi)=2e\dot{N} = \frac{2\pi}{\Phi_0}
\frac{\partial V(\phi)}{\partial \phi}$.  
Here $N$ $(=-i\partial/\partial \phi)$ is the number operator
of Cooper pairs satisfying $[\phi, N] = i$.
The ground state of the junction is determined by the local minima
of the potential energy.
We find that
as $V_g$ varies,
the sign of the Josephson current changes,
i.e., there occurs the ground-state transition of the junction between the
$0$ state $|0\rangle$, and the $\pi$ state $|\pi\rangle$.
This behavior results from the facts that
${k^{(F)}_F}=\sqrt{2m(E^{(F)}_F-eV_g)/\hbar^2}$ is tuned by $V_g$ and that
the change of $k_F^{(F)}$ has the same effect as that of
the length $d$, since the 0-$\pi$ transition is controlled by the single variable
$h_0 d/k^{(F)}_F$. \cite{Radovic}
This gate-voltage operation has better controllability
of the transition than the previous experimental ways where
samples with different $d$ were used \cite{Kontos} or
temperature $T$ was tuned \cite{Sellier}.
%

The parameters of a typical InAs 2DEG are used for the calculation.
We use the Fermi energy~\cite{Wildoer}
$E^{(F)}_F=$ 10~meV and
the effective g-factor~\cite{Bjork} $g=-10$.
The magnetic field induced by a ferromagnetic insulator~\cite{Vancura} is about
$1 - 2$ T.
We choose 2~T for it, which gives $h_0\approx$
0.25~meV, i.e., $h_0/E^{(F)} \approx 0.025$.
We choose
the length of the 2DEG as $d\approx 0.2$~$\mu$m, which is shorter
than the elastic mean free path ($\sim 1$~$\mu$m)~\cite{Takayanagi}
of electrons in the 2DEG, so we ignore disorders.
Under these
parameters, the ground-state transition occurs around
$eV_g/E^{(F)}_F \simeq -0.04$, where the two states, $|0\rangle$ and
$|\pi \rangle$, are degenerate; see Fig.~\ref{CPR}.
Note that a ferromagnetic insulator inducing weaker $h_0$ can cause
the same $0$-$\pi$ transition for a 2DEG with larger $d$,
as the transition is governed by the variable $h_0 d/k^{(F)}_F$.

So far, the charging energy of the insulating barriers (I) has been
ignored.  We assume that it negligibly modifies $V(\phi)$. Then the
Hamiltonian of the SIFIS junction is
\begin{equation}
H= -4 E_C\frac{\partial^2}{\partial \phi^2} + V({\phi}),
\label{H}
\end{equation}
where $E_C \equiv e^2/2C$ and $C$ is the junction capacitance.
The first term is the charging energy 
and causes the tunneling between $|0\rangle$ and $|\pi\rangle$.
The tunneling gives rise to the quantum mechanical superposition of the two states
when the two states are degenerate.
The tunneling rate is estimated by using the WKB approximation,
$t_\textrm{q} \simeq (\hbar\omega/2\pi)
 \exp {\left[- \sqrt{2M/\hbar^2}
    \int d\phi \sqrt{V(\phi)-E_{g}} \right]}$,
where $M=\hbar^2/8E_C$, $M \omega^2=
\partial^2V(\phi)/\partial\phi^2$, and $E_g=\hbar\omega /2$ in the
harmonic potential approximation. For smaller $C$, $t_\textrm{q}$
is larger. In usual superconducting qubits \cite{ChargeQ,Mooij},
$C \sim 10^{-15}$~F.
The capacitance of superconductor/2DEG junction depends on the
junction area, $C\sim 10 {\rm fF}/\mu{\rm m}^2$. \cite{Krasnov}
Here we set $C =$ 1~fF, and estimate
$t_\textrm{q}/h \approx 14$~GHz.

The $0$ and $\pi$ states constitute a qubit.
The qubit state is described by the superposition of $|0
\rangle$ and $|\pi \rangle$,
and it is operated by  the gate voltage $V_g$ applied to the
ferromagnetic 2DEG at the degeneracy point.
When
an oscillating gate voltage $\delta
V_g \cos \omega t$ with $e\delta V_g \ll E^{(F)}_F$ is applied,
the qubit state is described by the Hamiltonian,
\begin{eqnarray}
H_{\rm qubit}&=&E_0|0\rangle\langle 0|+E_\pi|\pi\rangle\langle \pi|
-t_\textrm{q}(|0\rangle\langle \pi|+|\pi\rangle\langle 0|) \nonumber\\
&+&g\cos\omega t (|0\rangle\langle 0|-|\pi\rangle\langle \pi|),
\end{eqnarray}
since $E_0$ and $E_\pi$ linearly depend on $V_g$ around the
degeneracy point.
The Hamiltonian  shows the Rabi oscillation between
$|S\rangle = ( |0\rangle +|\pi\rangle)/\sqrt{2}$ and
$|A\rangle = ( |0\rangle - |\pi\rangle)/\sqrt{2}$
with the period of  $2\pi\hbar/g$, when $\omega$ is identical to
the energy gap $2t_q$ at the degeneracy point, $E_0= E_\pi$. \cite{Ashhab}
The coupling strength $g \propto \delta V_g$ is determined by
junction parameters, and estimated as $g/2\pi\hbar\approx 1.5$\,GHz
for $\delta V_g \approx5$\,$\mu V$.
This scheme has the merit that  the operations take place at the optimal point, $\partial
E_{S(A)}/\partial V_g = 0$.

\begin{figure}[t]
\vspace{4.7cm}
\includegraphics{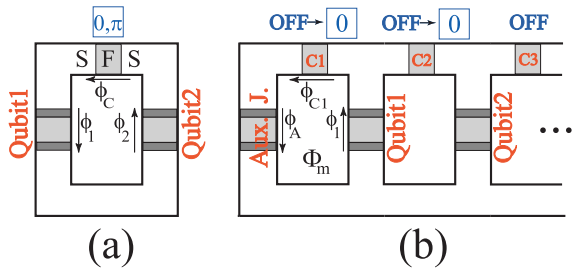}
\vspace{-0.7cm}
\caption{ (Color online) (a) Two coupled SIFIS
qubits.  The phase difference across Qubit $i=1,2$ (the SFS
junction) is denoted as $\phi_i$ ($\phi_\textrm{C}$). (b) A scalable scheme
for the measurement and coupling  of the SIFIS qubits.}
\label{coup}
\end{figure}

For a quantum computing the coupling of the qubits needs to be
controllable and switchable. We propose a scalable scheme for
qubit readout and tunable qubit coupling.
In Fig.~\ref{coup}(a), two SIFIS
and one SFS junctions form a loop, and the two SIFIS phase qubits,
saying Qubit1 and Qubit2, are coupled due to the periodic boundary
condition, $\phi_1+\phi_2+\phi_\textrm{C}=2\pi n$. The resulting
two-qubit state depends on the gate voltage applied to the SFS
junction. When the gate voltage is adjusted such that the SFS
junction is in the 0 ($\pi$) state, i.e., $\phi_C = 0$ ($\pi$),
Qubit1 and Qubit2 are in a superposition of $|0 0 \rangle$ and $|\pi
\pi \rangle$ ($|0 \pi \rangle$ and $|\pi 0 \rangle$). On the other
hand, when the gate voltage is adjusted such that both the 0 and
$\pi$ states are stable in the SFS junction, but need not be
degenerate, the energy difference between the two SFS states causes
the energy difference between the two-qubit states with
$\phi_\textrm{C} =0$ ($|00\rangle$ and $|\pi\pi\rangle$) and those
with $\pi$ ($|0\pi\rangle$ and $|\pi 0\rangle$).

This coupling scheme has an important advantage that the coupling
between qubits is tunable by the gate voltage applied on the SFS junction.
The coupling energy $J$ comes from the energy difference between the
states with $\phi_\textrm{C} = 0$ ($|0 0\rangle$ and $|\pi
\pi\rangle$) and those with $\phi_\textrm{C} = \pi$ ($|\pi 0\rangle$
and $|0\pi\rangle$). As the energy difference is tuned by the gate
voltage applied to the SFS junction, one can change the sign as well
as the strength of $J$. When $J=0$, the coupling is switched off
effectively.

After  two-qubit operations, the couplings
between the qubits in Fig.~\ref{coup}(b) are switched off.
Then the state of Qubit1 can be measured by using
the left auxiliary SIFIS junction, as follows.
One sets the gate voltages on C1 and on the auxiliary junction
such that only the 0 state is stable in both the junctions, i.e.,
the $\pi$ state cannot be formed even at the excited levels. Then one
turns on a small measuring external flux $\Phi_\textrm{m}$  through
the loop consisting of C1, Qubit1, and the auxiliary junction.
Then, the direction of current $I_0$ along
the loop is different for two states,  $|0 \rangle$ and $| \pi \rangle$,
of Qubit1. \cite{Yamashita06} Hence one
measures Qubit1 by observing the direction of induced magnetic moment
using a dc-SQUID \cite{Mooij}.
After the measurement of Qubit1, one can also detect Qubit2 by
setting the gate voltages such that the coupling junctions C1 and C3
are switched off and that Qubit1 and C2 are in the 0 states. Here
Qubit1 behaves as an auxiliary junction for Qubit2. In this way, all
the qubits in the whole circuit can be measured sequentially.


In summary, we have proposed a new qubit in an SIFIS junction,
and discussed a scalable scheme for qubit operation, measurement,
and tunable and switchable two-qubit coupling. The qubit is based on
the phase degree of freedom, and a single-qubit state and the
two-qubit coupling are operated by gate voltage. Our
scheme has an important merit that it is insensitive to the
decoherence coming from the fluctuations of background charges and
from magnetic microwave-pulse operations.

This work is supported by KRF (2005-070-C00055, 2006-331-C00118; TWN
and HSS).

\end{document}